\documentclass[a4paper]{PoS}

\newcommand{\be}{\begin{equation}}
\newcommand{\ee}{\end{equation}}
\newcommand{\bea}{\begin{eqnarray}}
\newcommand{\eea}{\end{eqnarray}}

\newcommand{\MKK}{M_{\rm KK}}

\newcommand{\uKK}{u_{\rm KK}}

\def\PDG{\cite{PDG16}}

\def\BRb{\cite{1510.07605}}
\def\PLB{\cite{Brunner:2016ygk}}
\def\WSS{Witten-Sakai-Sugimoto}

\title{Predictions for production and decay of\\ the pseudoscalar glueball from the Witten-Sakai-Sugimoto model}

\ShortTitle{Pseudoscalar glueball} % from the Witten-Sakai-Sugimoto model}

\author{Frederic Br\"unner and \speaker{Anton Rebhan}\\%\thanks{A footnote may follow.}\\
       Institute f\"ur Theoretische Physik, Technische Universit\"at Wien,\\ Wiedner Hauptstra{\ss}e 8-10, A-1040 Vienna, Austria\\
        E-mail: \email{bruenner,rebhana @hep.itp.tuwien.ac.at}}
\abstract{The Witten-Sakai-Sugimoto model, a top-down gauge/gravity model of large-$N_c$ low-energy QCD based on type-IIA string theory and $N_c$ D4 branes
with chiral quarks added by probe D8 branes, involves only one dimensionless parameter, the 't Hooft coupling at its cutoff (Kaluza-Klein) scale. Although this cutoff
scale is around 1 GeV, the model has turned out to be surprisingly predictive 
also quantitatively, reproducing masses of vector and axial vector mesons,
their decay rates,
as well as the anomalous mass of the $\eta'$ meson, all within 10--30\% errors. Using it as a guide for glueball signatures,
we have argued that it indicates that the meson $f_0(1710)$ may be a nearly pure glueball, with rather specific predictions for 
the still-to-be-measured $4\pi$ and $\eta\eta'$
decays. Here we present our new predictions for the decay pattern of the (very narrow) pseudoscalar glueball, which is closely related to
the U(1)$_A$ problem that the model is in fact handling correctly.
}

\FullConference{The European Physical Society Conference on High Energy Physics\\
		5-12 July, 2017\\
		Venice}

\begin{document}
%4 pages
\section{Introduction}

The spectrum of (bare, unmixed) glueballs is rather well known from lattice QCD \cite{Morningstar:1999rf%,Chen:2005mg
}, but their interactions
and thus questions regarding their decay pattern or mixing with $q\bar q$ states are still wide open.
Correspondingly, there is no conclusive identification of any glueball in the meson spectrum.
Various phenomenological models identify alternatingly the mesons $f_0(1500)$ and $f_0(1710)$ as
dominantly glue (typically 50-70\% or 75-90\%, respectively), with some recent tendency towards the
latter possibility of a comparatively pure glueball nature of $f_0(1710)$ \cite{Cheng:2015iaa}. For the tensor glueball,
predicted by lattice at around 2.4 GeV, the situation is even more unclear.

The pseudoscalar glueball, which in quenched lattice QCD is found at 2.6 GeV, is particularly elusive.
In 1980, the isoscalar pseudoscalar $\iota(1440)$ was considered as the first glueball candidate in the
meson spectrum, which is now listed as two states, $\eta(1405)$ and $\eta(1475)$ by the Particle Data Group \PDG.
Together with $\eta(1295)$ there seems to be a supernumerary state beyond the first radial excitations
of the $\eta$ and $\eta'$ mesons, with $\eta(1405)$ singled out as a glueball candidate. However, this
is not supported by lattice QCD, also not by recent studies of unquenching effects \cite{Sun:2017ipk}. Moreover, the possibility
that $\eta(1405)$ and $\eta(1475)$ could after all be just one state $\eta(1440)$ is still under discussion \cite{Crede:2008vw}.

\section{Glueballs in the \WSS\ model}

The closest (top-down) holographic model of (large-$N_c$) QCD is currently provided by the \WSS\ model \cite{%Witten:1998zw,
Sakai:2004cn%,Sakai:2005yt
}.
Since this model has been quite successful in reproducing essentially all qualitative features of low-energy QCD,
and even performed surprisingly well when it comes to quantitative evaluations, it is certainly interesting
to explore it with regard to predictions for glueball decay rates and branching ratios.
This has been pioneered by Hashimoto et al.\ \cite{Hashimoto:2007ze} and revisited by the present authors together with Parganlija
\cite{Brunner:2015oqa,1504.05815,1510.07605,Brunner:2016ygk}.

The glueball spectrum has been worked out in Refs.~\cite{Constable:1999gb%,Brower:2000rp
}, where scalar and tensor glueballs correspond to
modes with equal masses in the higher-dimensional dilaton and graviton fields. An extra, much lighter mode is
provided by an ``exotic polarization'' along the artificial Kaluza-Klein dimension of the model. In Refs.~\cite{Brunner:2015oqa,1504.05815,1510.07605}
we have found that the decay pattern of this mode, which comes out at a mere 855 MeV, does not match any scalar glueball candidate,
unless there is strong mixing (which would make the \WSS\ model inapplicable). Studying mass deformations of
the model, we have found however in Ref.~\cite{1510.07605} that the decay pattern of $f_0(1710)$ can be reproduced
almost perfectly, provided the not yet measured branching ratio for $\eta\eta'$ decays remains well below
the current upper limit obtained by WA102 \cite{Barberis:2000cd}.

In the \WSS\ model, the pseudoscalar glueball is represented by fluctuations of the Ramond-Ramond (RR) 1-form field that
is responsible for the Witten-Veneziano mass for singlet $\eta_0$ pseudoscalar mesons from the U(1)$_A$ anomaly
contributions of order $1/N_c$, yielding
\be\label{mWV2}
m_{0}^2=\frac{N_f}{27\pi^2 N_c}\lambda^2\MKK^2.
\ee
This arises from
\be
S_{C_1}=-\frac1{4\pi(2\pi l_s)^6}\int d^{10}x\sqrt{-g}|\,\tilde F_2|^2
\quad\mathrm{with}\quad
\tilde F_2=\frac{6\pi\uKK^3 \MKK^{-1}}{u^4}
\left(\theta+\frac{\sqrt{2N_f}}{f_\pi}\eta_0\right)du \wedge dx^4,
\ee
where $\theta$ is the QCD theta angle and
$
\eta_0(x)=\frac{f_\pi}{\sqrt{2N_f}}\int dz \,{\rm Tr} A_z(z,x).
$
(Here $u=0\ldots\infty$ is the holographic coordinate of the bulk geometry, and $z=-\infty\ldots\infty$ the holographic coordinate along the 
joined D8-$\overline{\rm D8}$ branes; $x^4$ denotes the compactified extra spatial dimension, $x^4\simeq x^4+2\pi/\MKK$, by which
supersymmetry and conformal invariance is broken at low energies.)

\begin{figure}
\includegraphics[width=0.49\textwidth]{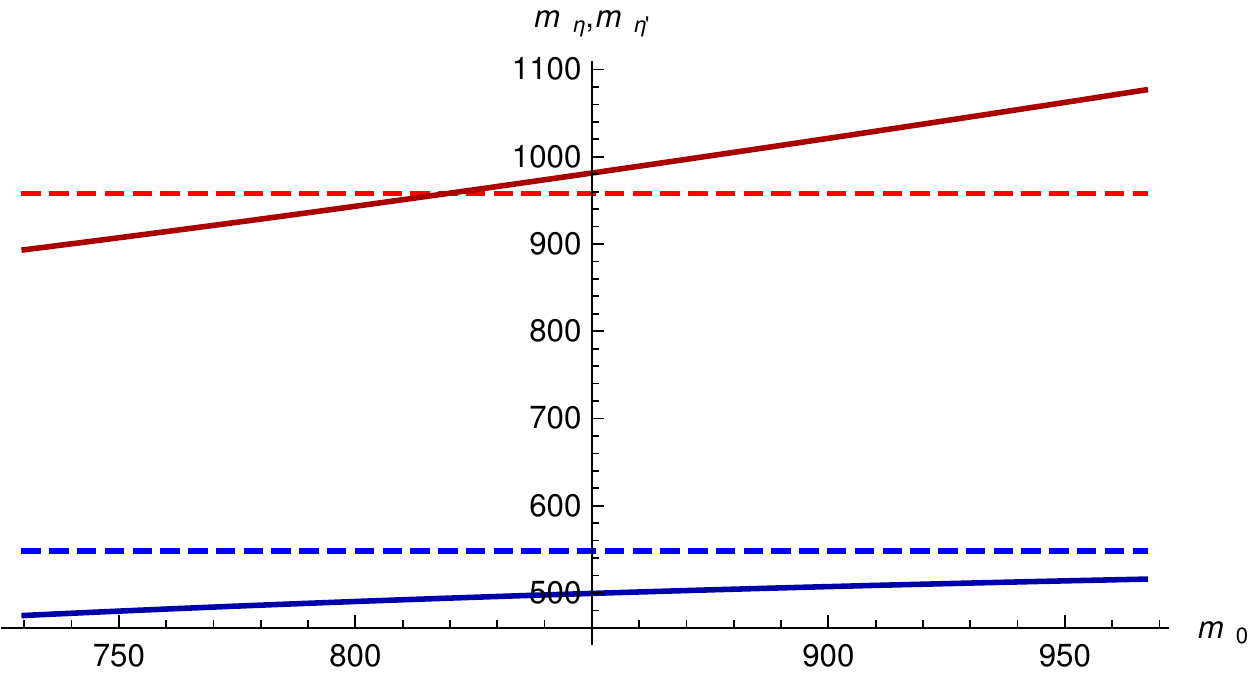}
\includegraphics[width=0.49\textwidth]{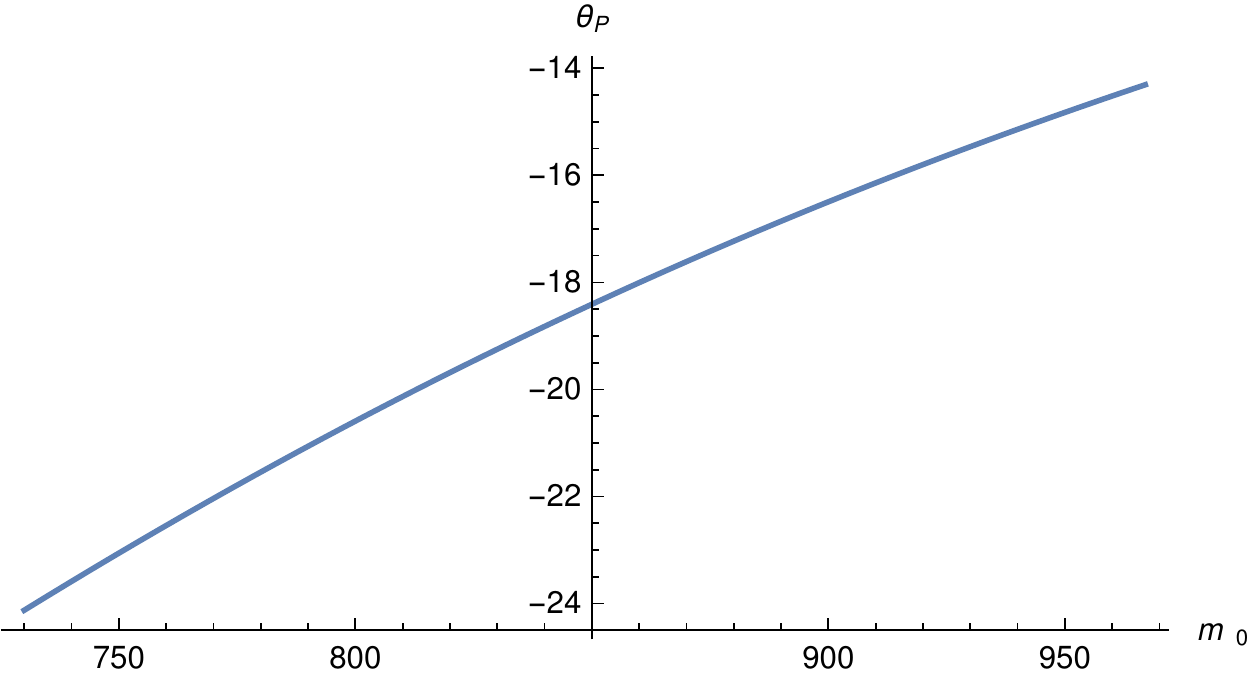}
\caption{Masses of the $\eta$ and $\eta'$ meson (blue and red lines in left plot, experimental values dashed) and pseudoscalar mixing angle $\theta_P$ (right plot)
as a function of the Witten-Veneziano mass $m_0$ in the range obtained for the latter in the Witten-Sakai-Sugimoto model. (Figure taken from \BRb.) }
\label{figmetas}
\end{figure}

For $N_f=N_c=3$, and the choice of parameters considered
by us in \cite{Brunner:2015oqa,1504.05815}, namely $\MKK=949$ MeV, and 't Hooft coupling $\lambda$ varied from 12.55 to 16.63, one finds
$m_{0}=730$\ldots$967$ MeV.
Including finite quark masses and rediagonalizing, one obtains masses for $\eta$ and $\eta'$ which agree to within 10\%
with the experimental values, and mixing angles that cover the range discussed in phenomenology (Fig.~\ref{figmetas}).

% \begin{figure}
% \includegraphics[width=.6\textwidth]{figure.eps}
% \caption{This is the caption of the figure.}
% \label{fig1}
% \end{figure}

The pseudoscalar glueball is described by fluctuations $C_1'$ in
\be\tilde F_2=
dC_1'+\frac{c}{u^4}%{3 \UKK^3 /U^4}{2\pi/\MKK}
\left( \theta+\frac{\sqrt{2N_f}}{f_\pi}\eta_0(x) \right)du \wedge dx^4.
\ee
Because there is no direct coupling of the RR field $C_1'$ to the flavor D8 branes, the \WSS\ model
does not lead to direct interactions of the pseudoscalar glueball with $q\bar q$ states.
The only relevant coupling is through a vertex involving a pseudoscalar glueball ($\tilde G$), the singlet $\eta_0$, and
a scalar glueball ($G$) -- without the latter, the potential mixing term of $\eta_0$ and
the pseudoscalar glueball field vanishes. The resulting $G$-$\tilde G$-$\eta_0$ vertex
turns out to read
$
2.583\ldots\,i\,\lambda^{1/2} N_f^{1/2} N_c^{-3/2}\MKK
$; those for $\eta$ and $\eta'$ are obtained by multiplication with $\sin\theta_P$ and $\cos\theta_P$, respectively.

\begin{figure}
\centerline{\includegraphics[width=0.5\textwidth]{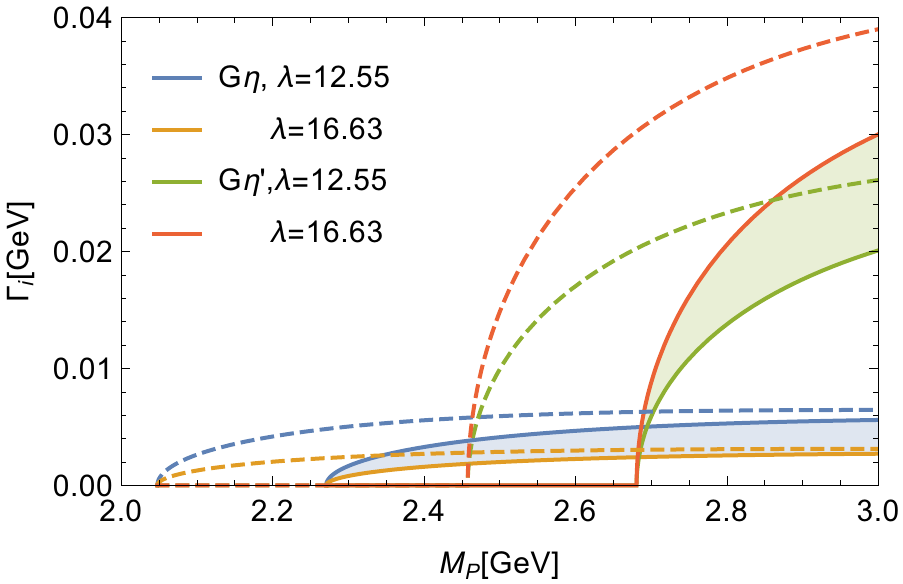}}
\caption{Partial width of the decay $\tilde{G}\to G\eta{(')}$ 
for a predominantly dilatonic scalar glueball $G_D$ with
mass $m_D=1.5$ GeV (dashed lines) and 1.723 GeV (full lines). (Figure taken from \PLB.)
}
\label{figdecayGTetaG}
\end{figure}

With $\MKK=949$ MeV, the (dilatonic) scalar glueball mass and the mass of the pseudoscalar glueball
are given by 1487 and 1789 MeV, respectively. For our extrapolation to the real world, we raise the value
of the scalar glueball mass to 1.723 GeV (corresponding to $f_0(1710)$) and the pseudoscalar glueball
mass to $2\ldots 3$ GeV. The resulting (resonant) decay rate is shown in Fig.~\ref{figdecayGTetaG} (neglecting the finite widths of the decay products).

Just as with decay, production of pseudoscalar glueballs to leading order involves a scalar glueball together with
$\eta(')$; the only other possibility is a vertex involving two pseudoscalar glueballs and a scalar glueball.
In fact, this would explain why no pseudoscalar glueball candidate has appeared yet in studies of radiative $J/\psi$ decays,
since the co-production threshold is larger than the mass of $J/\psi$, if the pseudoscalar glueball mass is around 2.6 GeV;
one would need excited $\psi$ or $\Upsilon$ instead.

\begin{figure}
\bigskip
\centerline{\hfil\includegraphics[width=0.1\textwidth]{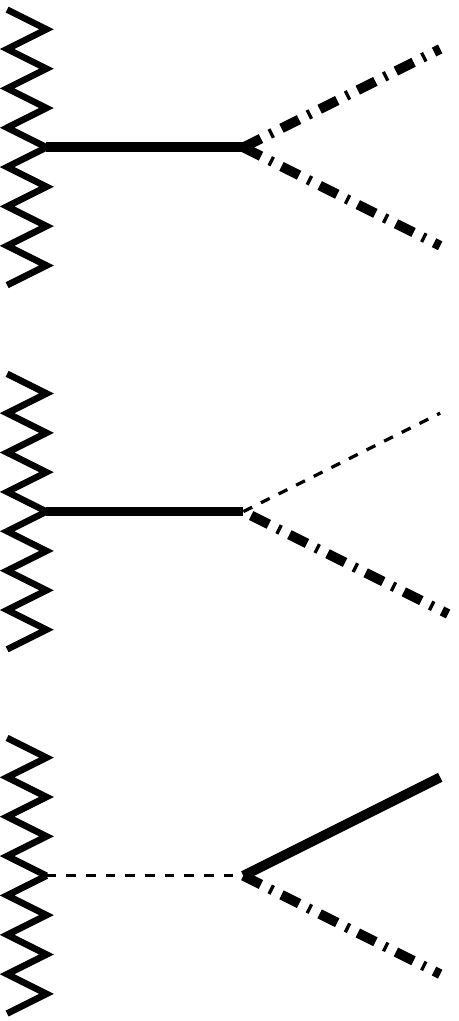}\hfil\hfil\hfil}
\begin{picture}(0,0)
\put(210,95){$\sim\lambda^{-1}N_c^{-2}$}
\put(210,60){$\sim\lambda^{0}N_f^{1/2}N_c^{-5/2}$}
\put(210,25){$\sim\lambda^{-1}N_f^{1}N_c^{-3}$}
\put(175,105){\scriptsize $\tilde G$}
\put(175,85){\scriptsize $\tilde G$}
\put(175,70){\scriptsize $\eta(')$}
\put(175,50){\scriptsize $\tilde G$}
\put(175,35){\scriptsize $G$}
\put(175,15){\scriptsize $\tilde G$}
\end{picture}
\caption{Parametric orders of the production amplitudes of pseudoscalar glueballs 
in double Pomeron or double Reggeon exchange. (In the uppermost diagram the full line
stands for scalar or tensor glueball, in the others it denotes only the scalar glueball.)}
\label{figCEP}
\end{figure}
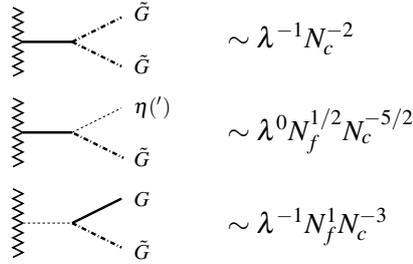

Another possibility is central exclusive production (CEP) in high-energy hadron collisions. Parametric orders for the various
production channels are given in Fig.~\ref{figCEP}.

\begin{figure}[t]
\centerline{\includegraphics[width=0.5\textwidth]{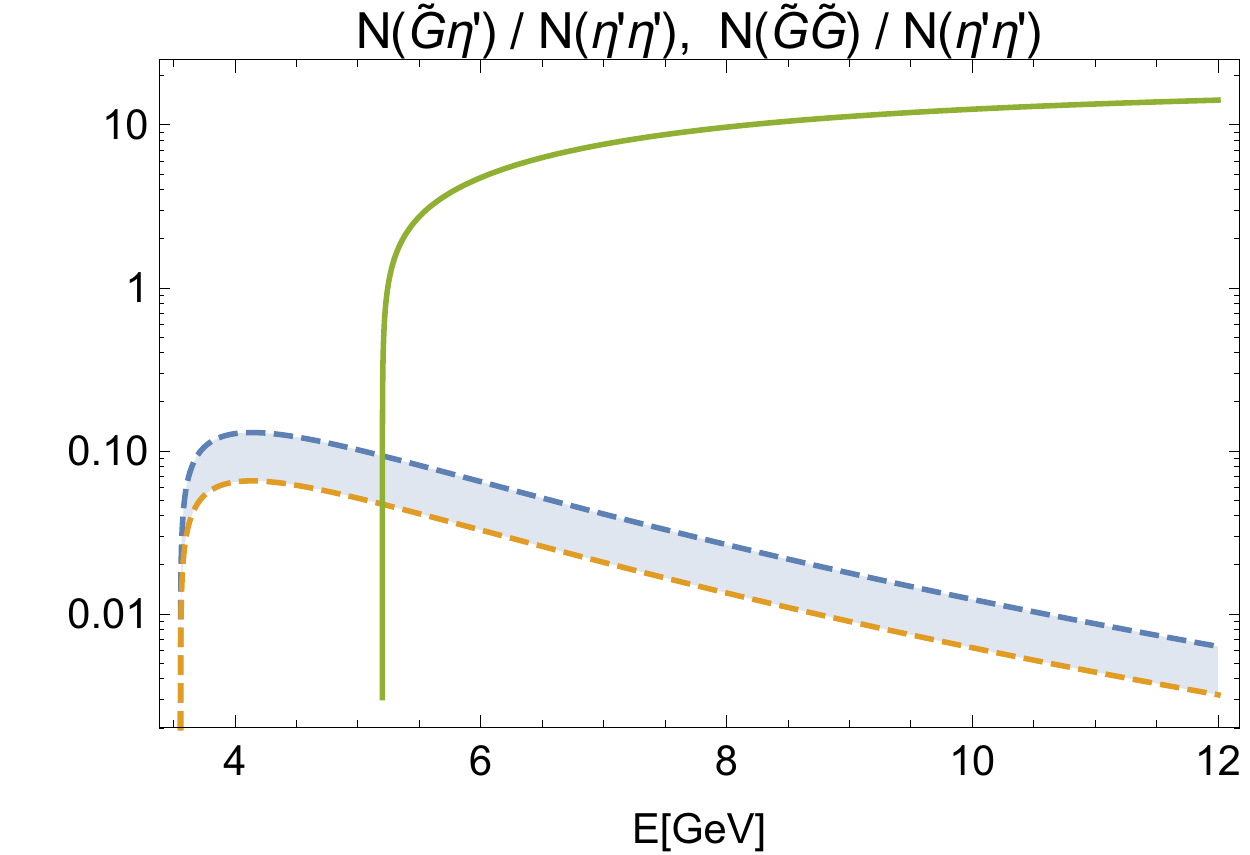}}
\caption{Production of $\tilde{G}\tilde{G}$ and $\tilde{G}\eta'$ pairs versus $\eta'\eta'$ from a virtual scalar glueball $G_D$ for
a pseudoscalar glueball mass of 2.6 GeV as functions of the center of mass energy of
the produced pair. The full line gives $N(\tilde{G}\tilde{G})/N(\eta'\eta')$,
which is independent of the 't Hooft coupling; upper and lower dashed lines
correspond to $N(\tilde{G}\eta')/N(\eta'\eta')$ with 't Hooft coupling 12.55 and 16.63, respectively. (Figure taken from \PLB.)}
\label{fig:pscgbproduction}
\end{figure}

In Fig.~\ref{fig:pscgbproduction} the production cross section of $\tilde{G}\tilde{G}$ and $\tilde{G}\eta'$ pairs versus $\eta'\eta'$ pairs
is given, assuming that all those arise from excited glue in the form of a virtual scalar glueball.
Assuming a pseudoscalar glueball mass of 2.6 GeV, pair production of two pseudoscalar glueballs turns out to be larger than $\eta'\eta'$
above threshold for the former, whereas $\tilde{G}\eta'$ pairs are one to two orders of magnitude lower.
As a reference point we note that CEP of $\eta'\eta'$ 
%in Durham [Harland-Lang et al. 2013]:
has been estimated in the Durham model as
$\sigma(\eta'\eta')/\sigma(\pi^0\pi^0)\sim 10^3\ldots10^5$  at $\sqrt{s}=1.96$~TeV \cite{Harland-Lang:2013ncy}.

These results seem to indicate that central exclusive production might be a promising way to finally observe the pseudoscalar glueball,
which according to the \WSS\ model should appear as a very narrow state with dominant decay into $\eta(')$ and a scalar glueball.

\begin{acknowledgments}
This work was supported by the Austrian Science
Fund FWF, project no. P26366, and the FWF doctoral program
Particles \& Interactions, project no. W1252.
\end{acknowledgments}

% \begin{thebibliography}{99}
% \bibitem{...}
% ....
% 
% \end{thebibliography}
\bibliographystyle{../JHEP}
%\bibliography{ar,tft,qft,books}
\bibliography{../glueballdecay}

\end{document}